\documentclass[16pt,a4paper]{article}
\usepackage{t1enc}
\usepackage[latin1]{inputenc}
\usepackage[english]{babel}
\long\def\symbolfootnote[#1]#2{\begingroup%
\def\thefootnote{\fnsymbol{footnote}}\footnote[#1]{#2}\endgroup} 

\title {Expressions for frictional and conservative force combinations within 
the dissipative Lagrange-Hamilton formalism}
\author{Charles E. Smith \footnote {\emph{E-mail address:}  cesmith\emph{@}vt.edu}
\\\footnotesize \emph{Dept. of Mechancial Engineering,} 
\\ \footnotesize \emph{Virginia Polytechnical Institute and State University, Blacksburg  VA, 24060, USA}}

\date{}

\begin{document}
\maketitle

\begin{abstract}

\leftskip=-1cm \rightskip=-1.5cm
\noindent Dissipative Lagrangians and Hamiltonians having Coulomb, viscous and quadratic damping,
together with gravitational and elastic terms are
presented for a formalism that preserves the Hamiltonian as a constant of the motion. 
Their derivations are also shown. The resulting $L$'s and  $H$'s may prove useful in exploring new types of 
damped quantum systems.

\vspace{2mm}

\noindent \footnotesize  \emph{Keywords:}  Dissipative Lagrangians; Generalized Hamilton's principle; Damped quantum systems  
\noindent \footnotesize  \emph{PACS numbers:} 45.20.Jj, 02.30.Xx, 02.30.Zz, 03.65.Ca
\end{abstract}

\vspace{3mm}


Of the Lagrange-Hamiltonian formalisms that incorporate frictional forces, 
the formulation by Schuch \cite{dS90}  is closely analagous to a conservative system.
This approach has had a number of advantages, especially for describing quantum systems.  
As expected, the Lagrangian $L$ produces the correct equation of motion.  The Hamiltonian $H$ is a 
constant of the motion. $H$ also has a physical interpretation of being the total energy for
the system.\symbolfootnote[2]{Depending on the initial conditions that are specified.}
The problem of shrinking Heisenberg uncertainty, often seen with canonical attempts at 
dissipative quantization  \cite{dS90,dS97,dS99,mR87}, 
can be resolved. In addition, the range of candidates for $H$ is greatly narrowed from an 
infinite set of possibilities, e.g. \cite{cY78,rS78}. For convenience, the formalism developed by 
Schuch will be referred to as  $\tilde{H}_{exp}$, where  \emph{exp} represents an 
expanding coordinate space. 

The existence of $\tilde{H}_{exp}$ does not necessarily rule out the usefulness of alternate forms. 
Schuch \cite{dS97,dS99} has also shown that well known formalisms that do not preserve
$H$ as a constant of the motion such as Caldirola \cite{pC41} and Kanai \cite{aK48}, can be converted
to and from $\tilde{H}_{exp}$ via canonical transformation.  There is a similar
relationship for the non-linear Schr\"{o}dinger equation, (NLSE), \cite{dS97,dS99}.
These other formalisms and $\tilde{H}_{exp}$ are closely related because  
they share similar \emph{non-canonical} transformations that connect the  physical and 
transformed coordinate spaces. 

Given its promising characteristics, new instances  are presented to show that
the $\tilde{H}_{exp}$ formalism can be broadened to include additional combinations of friction and potential. 
Some of the more important details concerning $\tilde{H}_{exp}$ are discussed briefly.

To obtain  $L$'s and $H$'s that have the desired properties, it is necessary to 
determine relations that connect the physical space coordinate $q$, to a transformed $Q$ space \cite{dS90}, 
\begin{equation}
Q = {\cal F} (q,t,\xi_0, \xi_1,\dots, \phi _0,\phi _1,\dots)
\end{equation}
where $t$ is time and $ \xi _i$ and $\phi _i$ are the coefficients of friction and potential.  Note that within $Q$-space explicit
time dependence is absent. 

The resulting $L$'s and $H$'s for a prospective system with frictional forces 
must also satisfy a set of formalisic rules \cite{dS90} which are listed below.  
  
\begin {enumerate}
\leftskip=-1cm \rightskip=-1.5cm

\item  All $L(\dot {q},q,t)$'s, $L(\dot{Q},Q)$'s,  $H(p,q,t)$'s, and $H(P,Q) $'s have the units of energy.
\item  The $L$ for the system has an invariance property whereby both 
$\delta L(\dot {q},q,t)=0$ and $\delta L(\dot{Q},Q)$ \hspace{1mm}$=0$ correctly produce the physical space equation of motion. 
\symbolfootnote[3]{$\delta L$ represents the variation of the action integral $\delta \int_{t_1}^{t_2} L dt $ and is
calculated from the Euler-Lagrange relation $\frac{d}{dt} \frac{\partial L}{\partial \dot x} - \frac{\partial L}{\partial x} = 0$.} 
\item The physical and transformed momenta $p$ and $P$ are defined as $p = m\dot q$ and $P = m\dot Q$ respectively. In the 
absence of expressed dissipative forces, as in $Q$-space, $P  =  \frac{\partial L(\dot{Q},Q)}{\partial \dot Q}$ also holds. 
If friction is expressed explicitly, as in real space, $p \not=  \frac{\partial L(\dot{q},q, t)}{\partial \dot q }$ in general!
\item  $ H(P,Q) $ is found via the Lengendre transform $ H= P\dot{Q}-L$ where neither
$L= L(\dot{Q},Q) $ nor $ H(P,Q) $ contain $t$ explicitly. If there is explicit time dependence in the connection
between $Q$ and $q$, $H(p,q,t)\not=p\dot{q}-L(\dot{q},q,t)$! \symbolfootnote[4]{The need for condition 4
is seen for instance, whenever viscous friction is present.  For quadratic friction, where time does 
not appear explicitly, $H$ can indeed be found by using the Legendre transform in $q$-space.}
\item The $H(p,q,t)$ that is obtained from $ H(P,Q) $ is a constant of the motion and can be made equal to the initial 
energy.
\item As the coefficients for friction or potentials are made to vanish, $Q$,  $\dot{Q}$, $P$, $L(\dot {q},q,t)$, 
$L(\dot{Q},Q)$,  $H(p,q,t)$ and  $ H(P,Q) $ collapse smoothly to the expected simpler forms.
\item $Q$, $\dot{Q}$ as well as $q$, $\dot{q}$ are related through simple differentiation with respect to time:  
$\dot{Q} = \frac{dQ}{dt}$; $\dot{q} = \frac{dq}{dt}$.
\
\end {enumerate}

Previously derived examples for $\tilde{H}_{exp}$ include the damped free particle and the damped harmonic 
oscillator \cite{dS90}.   
For a damped oscillator, with $\ddot{q}+\gamma\dot{q}+\omega^2q=0$, the $Q$, $L$'s, $H$'s 
are:

\vspace{2mm}
\indent $Q = q\exp(\frac{\gamma t}{2})$

\indent $L(\dot{Q},Q)$ = $\frac{m}{2}(\dot{Q}^2 - (\omega^2-\frac{\gamma^2}{4})Q^2)$

\indent $H(P,Q)$ = $\frac{m}{2}(\frac{1}{m^2} P^2 + (\omega^2-\frac{\gamma^2}{4})Q^2)$

\indent $L(\dot{q},q,t)  = \frac{m}{2} (\dot{q}^2  + \gamma \dot{q}q + \frac{\gamma ^2}{2}q^2 -\omega ^2q^2) \exp(\gamma t)$

 \indent $H(p,q,t)  = \frac{m}{2} (\frac{1}{m^2} p^2  + \frac{ \gamma}{m} pq  +\omega ^2q^2) \exp(\gamma t)$
\vspace{2mm}

\noindent Solving the associated equation of motion for $p(t)$ and $q(t)$ and inserting the results back into $H(p,q,t)$,
shows that $H(p,q,t)$ is a constant of the motion.   $H(p,q,t)$ is also equal to the initial energy under the right inital conditions,
such as $q_0=0$ or $\dot q_0=0$ for example.  

$Q$'s, $L$'s and $H$'s that satisfy conditions 1-7 for a particle moving in a constant gravity field or against friction for
the lowest powers of velocity are shown in tables 1-a and 1-b. 

\vspace{2mm}
\noindent Table 1-a.
\vglue 1mm

\noindent \begin{tabular}{|l|l|l|l|}
   \hline
   eqn of motion & $Q$ definition & $L(\dot{Q},Q)$ & $H(P,Q)$ \\
   \hline
   $\ddot{q}+g=0$ & $Q=q$   & $\frac{m}{2}(\dot{Q}^2 -2gQ)$ & $\frac{m}{2}(\frac{1}{m^2}P^2 +2gQ)$  \\
   $\ddot{q}+\mu=0$ & $Q=q\exp(0)$   & $\frac{m}{2}(\dot{Q}^2 -2\mu Q)$ & $\frac{m}{2}(\frac{1}{m^2}P^2 +2\mu Q)$  \\
   $\ddot{q}+\gamma\dot{q}=0$ & $Q=q\exp(\frac{\gamma}{2}t)$ &$\frac{m}{2}(\dot{Q}^2+\frac{\gamma^2}{4}Q^2)$ & $\frac{m}{2}(\frac{1}{m^2}P^2-\frac{\gamma^2}{4}Q^2)$ \\
    $\ddot{q}+b \dot{q}^2=0$ & $Q=\frac{1}{b} (1-\exp(-bq))\exp(bq)$ & $\frac{m}{2}\dot{Q}^2$ & $\frac{1}{2m}P^2$\\
   \hline
   \end{tabular}
\vskip 1mm
\vspace{2 mm}
\noindent Table 1-b.
\vglue 1mm

\noindent \begin{tabular}{|l|l|l|l|}
   \hline
   eqn of motion  & $L(\dot{q},q,t)$  & $H(p,q,t)$ \\
   \hline
   $\ddot{q}+g=0$        &  $\frac{m}{2}(\dot{q}^2 -2gq)$                                                                                                                      & $\frac{m}{2}(\frac{1}{m^2}p^2 +2gq)$\\
   $\ddot{q}+\mu=0$        &  $\frac{m}{2}(\dot{q}^2 -2\mu q)\exp(0)$                                                                                                          & $\frac{m}{2}(\frac{1}{m^2}p^2 +2\mu q)\exp(0)$\\
      $\ddot{q}+\gamma\dot{q}=0$  &  $\frac{m}{2}(\dot{q}^2 +\gamma\dot{q}q+\frac{\gamma^2}{2}q^2)\exp(\gamma t)$   & $\frac{m}{2}(\frac{1}{m^2}p^2 +\frac{\gamma}{m}pq)\exp(\gamma t)$ \\
    $\ddot{q}+b \dot{q}^2=0$          & $\frac{m}{2}\dot{q}^2\exp(2bq)$                                                                                                 & $\frac{1}{2m}p^2 \exp(2bq)$ \\
   \hline
   \end{tabular}
\vspace{3mm}

The entries of table 1 have additional noteworthy characteristics.  Each $q$-space $L$ and $H$ contains 
a dimensionless expansion 
factor\symbolfootnote[5]{Also known as an integrating factor.} as indicated by Yan \cite{cY78}: 
\begin{equation}
    X=exp [-\int \frac{\partial G}{\partial \dot q} dt] 
\end{equation}
where $G$ is the right hand part of the generic differential equation, $\ddot{q}=G(\dot q,q,t)$.
Conservative potentials that exist in physical space are required by condition 6 to be explicity 
expressed within $ L(\dot{Q},Q) $ and $ H(P,Q) $ as well.  Some cases, like the viscous example, require
potential-like terms such as $\frac{\gamma^2}{4}Q^2$ in $ L(\dot{Q},Q) $ and $ H(P,Q) $  for the conditions 
of $\tilde{H}_{exp}$ to be fulfilled.  However, for the quadratically damped case,
no such terms are needed. As will be seen, other frictional systems also require "pseudopotentials" that 
have the form $f(Q,\xi_0, \xi_1,\dots, \phi _0,\phi _1,\dots)$ where $Q$ is the transformed position and $ \xi _i$ and $\phi _i$ are 
the relevant coefficients of friction and potential.   

The tables 2-a and 2-b list the $Q$'s, $L$'s and $H$'s derived for various combinations of friction and potential. 
The symbol $\alpha$ is used to represent gravity $g$ or  dry friction $\mu$ that opposes the motion.
Note that each $L(\dot{Q},Q)$ and $H(P,Q)$ contains some form of pseudopotental.

\vspace{2mm}
\noindent Table 2-a.
\vglue 1mm

\noindent \begin{tabular}{|l|l|l|l|l|}
   \hline
   eqn of motion & $Q$ definition & $L(\dot{Q},Q)$ & $H(P,Q)$ \\
   \hline
   $\ddot{q}+\gamma\dot{q}+\alpha=0$ & $Q=(q+\frac{\alpha t}{\gamma}-2\frac{\alpha}{\gamma^2})\exp(\frac{\gamma t}{2})$  & $\frac{m}{2}(\dot{Q}^2 +\frac{\gamma^2}{4}Q^2-2\alpha Q)$ & $\frac{m}{2}(\frac{1}{m^2}P^2 -\frac{\gamma^2}{4}Q^2+2\alpha Q)$  \\
                                                               &\indent $-2\frac{\alpha}{\gamma^2}\cosh(\frac{\gamma t}{2})+4\frac{\alpha}{\gamma^2}$ &&\\
                                                              &&&\\

   $\ddot{q}+b\dot{q}^2+\gamma\dot{q}=0$ & $Q=\frac{1}{b} (1-\exp(-bq))$                                                  & $\frac{m}{2}(\dot{Q}^2 +\frac{\gamma^2}{4}Q^2)$ &  $\frac{m}{2}(\frac{1}{m^2}P^2 -\frac{\gamma^2}{4}Q^2)$\\
                                                                                &   \hspace{15mm}$\times\exp(bq+\frac{\gamma}{2}t)$    &&\\
                                                                      & &&\\

   $\ddot{q}+b\dot{q}^2+\alpha =0$              & $Q=\frac{1}{b} (1-\exp(-bq))\exp(bq)$                                              & $\frac{m}{2}(\dot{Q}^2 -\alpha bQ^2 - 2\alpha Q)$ &  $\frac{m}{2}(\frac{1}{m^2}P^2 +\alpha bQ^2 + 2\alpha Q)$\\
                                                                     & &&\\                                                                   
                                                                 
   $\ddot{q}+b\dot{q}^2+\omega^2q =0$  & $Q=\frac{1}{b} (1-\exp(-bq))\exp(bq)$                                           & $\frac{m}{2}(\dot{Q}^2$                                                                 &  $   \frac{m}{2}(\frac{1}{m^2}P^2 $\\
                                                                        &                                                                                                                 & $-\frac{\omega^2}{b^2}\ln(bQ+1)(bQ+1)^2$                                 &   $ +\frac{\omega^2}{b^2}\ln(bQ+1)(bQ+1)^2$\\
                                                                        &                                                                                                                 & $+\frac{\omega^2}{2b^2}(bQ+1)^2 -\frac{\omega^2}{2b^2} )$ &   $  -\frac{\omega^2}{2b^2}(bQ+1)^2 +\frac{\omega^2}{2b^2} )$\\
                                                                     & &&\\                                                                   
                                                      
      \hline
 \end{tabular}

\vspace{2mm}
\noindent Table 2-b.
\vglue 1mm

\noindent \begin{tabular}{|l|l|l|l|l|}
   \hline
   eqn of motion & $L(\dot{q},q,t)$  & $H(p,q,t)$ \\
   \hline
   $\ddot{q}+\gamma\dot{q}+\alpha=0$  &  $\frac{m}{2}((\dot{q}^2 +\gamma\dot{q}q+\frac{\gamma^2}{2}q^2$                                                      &  $\frac{m}{2}((\frac{1}{m^2}p^2 + \frac{\gamma}{m} pq  $\\
                                                                &  $ -2\alpha q- \frac{\alpha}{\gamma}\dot{q}+2\frac{\alpha^2}{\gamma^2}  -  2\frac{\alpha^2}{\gamma}t + \alpha\dot{q}t $    &  $ + \alpha q - \frac{\alpha}{m\gamma} p -2\frac{\alpha^2}{\gamma^2}+ \frac{\alpha^2t}{\gamma}+ \frac{\alpha}{m}pt )\exp(\gamma t)$ \\
                                                                 &$ +\alpha\gamma q t + \frac{\alpha^2t^2}{2} )\exp(\gamma t)$                                                                                    &  $+ \alpha q +\frac{\alpha}{m\gamma}p + 2\frac{\alpha^2}{\gamma^2} + \frac{\alpha^2t}{\gamma})$\\
                                                                &$ + \frac{\alpha}{\gamma}\dot{q} - 3 \frac{\alpha^2}{\gamma^2}+ \frac{\alpha^2}{\gamma^2}\cosh(\gamma t))$   &\\
                                                                    & &\\

   $\ddot{q}+b\dot{q}^2+\gamma\dot{q}=0$ & $\frac{m}{2}(\dot{q}^2 \exp(2bq+\gamma t) $                                        &  $   \frac{m}{2}(\frac{1}{m^2}p^2 \exp(2bq+\gamma t) $ \\
                                                                                &  $+ \frac{\gamma}{b}\dot{q}(\exp(bq)-1) \exp(bq+\gamma t) $        &   $+ \frac{\gamma}{mb}p \hspace{1mm}(\exp(bq)-1) \exp(bq+\gamma t)) $ \\
                                                                                & $+ \frac{\gamma^2}{2b^2}(\exp(bq)-1)^2\exp(\gamma t))$                   &            \\
                                                                                &&\\ 

   $\ddot{q}+b\dot{q}^2+\alpha=0$                      &  $\frac{m}{2}((\dot{q}^2-\frac{\alpha}{b})\exp(2bq)  + \frac{\alpha}{b})$               &$\frac{m}{2}((\frac{1}{m^2}p^2 + \frac{\alpha}{b})\exp(2bq) - \frac{\alpha}{b})$  \\                                                                           
                                                                            &&\\

    $\ddot{q}+b\dot{q}^2+\omega^2q=0$          &  $\frac{m}{2}((\dot{q}^2-\frac{\omega^2}{b}q+\frac{\omega^2}{2b^2})\exp(2bq)  - \frac{\omega^2}{2b^2})$      &$\frac{m}{2}((\frac{1}{m^2}p^2+\frac{\omega^2}{b}q-\frac{\omega^2}{2b^2})\exp(2bq)  + \frac{\omega^2}{2b^2}) $  \\                                                                           
                                                                            &&\\

        \hline
   \end{tabular}
\vspace{1mm}

\noindent It is readily demonstrated that both  $\delta L(\dot {q},q,t)=0$ and $\delta L(\dot{Q},Q)=0$  produce the correct 
equations of motion. Solving for $\dot{q}(t)$ and $q(t)$ and inserting the 
results back into $H(p,q,t)$, also shows that the $H(p,q,t)$'s are constants of the motion. Under requisite initial conditions,
 $H$ is equal to the initial energy as well.  Other combinations are also possible.

There are general characteristics of solutions that are observed. The combination 
of two or more $L(\dot{q},q,t)$'s with differing friction terms results in their respective 
expansion factors being multiplied together \cite{cY78,pH57,aT83}. The $Q$-space $L$ for a
combination includes the potentials and pseudopotentials of the simpler $L$'s.
For combinations of $\mu$ or $g$ with a form 
that has viscous friction, the $Q$-definition contains an additonal $f(t)$ that is to be determined.  

As has been mentioned, pseudopotential functions in $Q$-space are seen to be important in finding solutions to 
dissiptive $L$'s and $H$'s in the
Schuch formalism, but there is not yet a proved method for their determination.
The current examples follow the provisional rule that if such a function can be found
which satisfies conditions 1-7, then it is included.  In some cases, only one simple possiblity exists that 
has the correct units, i.e. energy.  For the free quadradically damped particle there evidently are none.
In the case of the quadratically damped oscillator, a pseudopotential may be determined by integrating the 
equation of motion $Q$-space.
Examples are presented to illustrate in more detail how the $Q$ definitions and subsequent $L$ and $H$ 
combinations were constructed.  

Example 1: $\ddot{q}+\gamma\dot{q}+\alpha=0$.
The ansatz $Q$-definition is taken to be:
\begin{equation}
     Q= q \exp(\frac{\gamma}{2}t) +  f(t)
\end{equation}
Since the invariance requirement of condition 2 means that the $L$ must produce the correct equation of motion 
regardless of whether it is in  $Q$ or $q$ space, $Q$-space may be used, which is more convenient.
An initial $L(\dot{Q},Q)$ is formed from the terms that correspond to each individual frictional
coefficient and potential as listed in table 1a:
\begin{equation}
     L(\dot{Q},Q)=\frac{m}{2}(\dot{Q}^2 +\frac{\gamma^2}{4}Q^2-2 \alpha Q)
\end{equation}
$\delta L(\dot{Q},Q)=0$ is calculated:
\begin{equation}
     \delta L(\dot{Q},Q)= \frac{m}{2}(2\ddot{Q} -\frac{\gamma^2}{2}Q+2 \alpha ) = 0
\end{equation}
The $Q$ above is differentiated to find  $\dot{Q}$ and $\ddot{Q}$, which are used to convert the differential 
equation to $q$-space:
The result is:
\begin{equation}
     (\ddot{q}+\gamma\dot{q})\exp(\frac{\gamma}{2}t)+\ddot{f}- \frac{\gamma^2}{4}f + \alpha=0
\end{equation}
The following differential equation in $f$ must therefore be solved:
\begin{equation}
     \ddot{f}- \frac{\gamma^2}{4}f +\alpha=\alpha\exp(\frac{\gamma}{2}t)
\end{equation}
The general solution for $f$ is of the form,
\begin{equation}
     f(t) = c_1 \exp(\frac{\gamma}{2}t) +c_2 \exp(\frac{-\gamma}{2}t) + \frac{\alpha t}{\gamma}\exp(\frac{\gamma }{2}t) + 4 \frac{\alpha}{\gamma^2}
\end{equation}
Using series expansions for $ c_1 \exp(\frac{\gamma}{2}t) $ and $c_2 \exp(\frac{-\gamma}{2}t)$, then applying condition 6, it is
found to be necessary that:
\begin{equation}
     c_1 = -3 \frac{\alpha}{\gamma^2}; \hspace{3mm} c_2= - \frac{\alpha}{\gamma^2}
\end{equation}
A valid $Q$ definition is therefore:
\begin{equation}
     Q = (q- 3 \frac{\alpha}{\gamma^2}+\frac{\alpha t}{\gamma} )\exp(\frac{\gamma}{2}t)  -  \frac{\alpha}{\gamma^2}\exp(\frac{-\gamma}{2}t) + 4 \frac{\alpha}{\gamma^2}
\end{equation}
After combining terms, $Q$ with resulting $L$'s and $H$'s are as shown in tables 2-a and  2-b.

Example 2: $\ddot{q}+b\dot{q}^2=0$.
Derivation of  the $Q$ definition can procede if one notices that $\frac{d}{dt}(\dot{q}\exp(bq))$ produces the desired 
equation of motion.
It is possible to set up a differential equation for an unknown function $f$ based on the assumptions:
\begin{equation}
     Q=f\exp(bq); \hspace{5mm}\dot{Q} = \dot{f}\exp(bq)+bf\dot{q}\exp(bq) = \dot{q}\exp(bq)
\end{equation}
The resulting differential equation is:
\begin{equation}
     \dot{f}+(bf-1)\dot{q}=0
\end{equation}
The solution for $f$ in terms of $q$ is then:
\begin{equation}
     f=\frac{1}{b} (1-\exp(-bq))
\end{equation}
A valid $Q$ definition is therefore:
\begin{equation}
     Q=\frac{1}{b} (1-\exp(-bq))\exp(bq)
\end{equation}
The method above works for quadratic friction,  but there is 
evidently no current technique for determining $Q$ definitions of orders $\dot{q}^3$ and higher.

The $L(\dot{q},q,t)$ that is obtained for Example 3 and shown in table 1-b is essentially the same as what is found in 
Havas \cite{pH57}, Tartaglia \cite{aT83} and Razavy~\cite{mR87}.  
Since time does not appear explicitly, the $L$'s still match even if potentials are
included \cite{aT83}.  However, $\tilde{H}_{exp}$ requires by conditions 3 and 5 that a system in real space be based on 
the kinetic momentum as opposed to the conjugate momentum.
Because the resulting $H(p,q,t)$  for this example uses $p = m\dot q$ instead of $p = \frac{\partial L}{\partial \dot q }$,
it differs from the $H$'s of the aforementioned authors.

Example 3: $\ddot{q}+b\dot{q}^2+\gamma\dot{q}=0$. 
The initial $Q$-definition is assumed to be the $Q$ for the highest order of friction is multipied by the lower order expansion factor:   
\begin{equation}
     Q=\frac{1}{b} (1-\exp(-bq))\exp(bq)\exp(\frac{\gamma}{2}t) = \frac{1}{b} (\exp(bq)-1)\exp(\frac{\gamma}{2}t)
\end{equation}
A combination $L(\dot{Q},Q)$ that includes the terms from each corresponding simpler $L$ is assumed:
\begin{equation}
L(\dot{Q},Q)=\frac{m}{2}(\dot{Q}^2 +\frac{\gamma^2}{4}Q^2)
\end{equation}
Then $\delta L(\dot{Q},Q)=0$ is calculated:
\begin{equation}
     \delta L(\dot{Q},Q)= \frac{m}{2}(2\ddot{Q} -\frac{\gamma^2}{2}Q) = 0
\end{equation}
Converting the differential equation to $q$-space obtains:
\begin{equation}
     (\ddot{q}+b\dot{q}^2+\gamma\dot{q})\exp(bq)\exp(\frac{\gamma}{2}t) =0
\end{equation}
The expansion factor is divided out and the desired result is confirmed with no other steps needed.

Example 4:  $\ddot{q}+b\dot{q}^2+\omega^2q=0$ 
The initial $Q$-definition is assumed to be:   
\begin{equation}
     Q=\frac{1}{b} (1-\exp(-bq))\exp(bq)
\end{equation}
The ansatz $L(\dot{Q},Q)$  should include $\omega^2Q^2$ but it will be more convenient to leave this term 
out for the time being:
\begin{equation}
L(\dot{Q},Q)=\frac{m}{2}(\dot{Q}^2 )
\end{equation}
The $\delta L(\dot{Q},Q)$ is calculated:
\begin{equation}
     \delta L(\dot{Q},Q)= \frac{m}{2}(2\ddot{Q}) = 0
\end{equation}
Converting the differential equation to $q$-space:
\begin{equation}
     (\ddot{q}+b\dot{q}^2)\exp(bq) =0
\end{equation}
It is clear that an $\omega^2q\exp(bq)$ term is required to complete the equation of motion. Even though friction is
present, there is no explicit time dependence. It is possible to work backward, converting the needed term to a 
function of $Q$ by using the $Q$-definition.
\begin{equation}
          \omega^2q\exp(bq) = \frac{\omega^2}{b}\ln(bQ+1)(bQ+1)
\end{equation}
The right hand side of equation (23) above is integrated to complete the desired $L(\dot{Q},Q)$ which results in
the complicated looking expression at the bottom of the third column in Table 2a.
By expanding these terms into a power series it is seen that  the conservative potential $\omega^2Q^2$ is present.
A constant $ -\frac{\omega^2}{2b^2}$ has also been added to satisfy condition 6. 
For $\ddot{q}+b\dot{q}^2+\alpha=0$  the same procedure as just described in this example can be used.

\vspace{5mm}
\noindent \textbf{Conclusion}
\vglue 5mm

It is possible to construct Lagrangians and Hamiltonians with additional combinations of friction and potential terms
that are consistent with a previously published Lagrange-Hamilton formalism  for dissipative systems. The resultant 
$L$'s and $H$'s offer new possibilities for canonical quantizations of frictional systems.  Other areas of future 
study include the
determination of a general method for finding Q definitions for higher order frictional cases.
The physical significance of the $L$'s and $H$'s, such as may exist, remains to be completely
explained.  A proof of uniqueness for $L$'s and $H$'s may require additional, as yet 
undefined conditions to ensure unique physical interpretations.


\end{document}